\documentclass[twocolumn]{article}
\usepackage{amsmath,amsthm}
\input epsf.sty

\newtheorem{dfn}{Definition}
\newtheorem{thm}{Theorem}

\newtheorem{cor}[thm]{Corollary}
\newtheorem{prob}{Problem}

\begin{document}

\newcommand{\ket}[1]{|{#1}\rangle}

\newcommand{\E}{\mathcal{E}}
\newcommand{\PG}{\mathcal{P}}

\newcommand{\la}{\leftarrow}

\title{The Minimum Distance Problem for Two-Way Entanglement
Purification%
\footnote{Personal use of this material is permitted.  However,
permission to reprint/republish this material for advertising or
promotional purposes or for creating new collective works for
resale or redistribution to servers or lists or to reuse any
copyrighted component of this work in other works must be obtained
from the IEEE.}}

\author{Andris Ambainis$^{1,2}$\thanks{E-mail: ambainis@iqc.ca}
and Daniel Gottesman$^3$%
\thanks{E-mail: dgottesman@perimeterinstitute.ca}\\
\\
${}^1$ School of Mathematics, Institute For Advanced Study, Princeton,
NJ 08540, USA\\
${}^2$ Institute for Quantum Computation, Waterloo, ON N2L 3G1
Canada\\
${}^3$ Perimeter Institute, Waterloo, ON N2V 1Z3 Canada}

\date{}

\maketitle

\begin{abstract}
Entanglement purification takes a number of noisy EPR pairs
$\ket{00} + \ket{11}$ and processes them to produce a smaller
number of more reliable pairs. If this is done with only a forward
classical side channel, the procedure is equivalent to using a
quantum error-correcting code (QECC).  We instead investigate
entanglement purification protocols with two-way classical side
channels (2-EPPs) for finite block sizes. In particular, we
consider the analog of the minimum distance problem for QECCs, and
show that 2-EPPs can exceed the quantum Hamming bound and the
quantum Singleton bound.  We also show that 2-EPPs can achieve the
rate $k/n = 1 - (t/n) \log_2 3 - h(t/n) - O(1/n)$ (asymptotically
reaching the quantum Hamming bound), where the EPP produces at
least $k$ good pairs out of $n$ total pairs with up to $t$
arbitrary errors, and $h(x) = -x \log_2 x - (1-x) \log_2 (1-x)$ is
the usual binary entropy. In contrast, the best known lower bound
on the rate of QECCs is the quantum Gilbert-Varshamov bound $k/n
\geq 1 - (2t/n) \log_2 3 - h(2t/n)$. Indeed, in some regimes, the
known upper bound on the asymptotic rate of good QECCs is strictly
below our lower bound on the achievable rate of 2-EPPs.
\end{abstract}


\section{Introduction}

In order to build a quantum computer, we will probably need to use
quantum error correcting codes (QECCs) to protect the
computational qubits from noisy operations (see~\cite{intro} for
an introduction to quantum error correction). Similarly, quantum
error correction will help preserve qubits stored in a quantum
memory. Another application is to protect quantum data being
transmitted over a distance from Alice to Bob.

For the last application, though, a better possibility exists.  In
an {\em entanglement purification protocol} (EPP)~\cite{BDSW},%
\footnote{The current trend is to instead use the name
``entanglement distillation protocol,'' or EDP.  However, in this
paper we retain the older and more widespread term EPP.}
Alice prepares a number of EPR pairs and transmits half of each to
Bob over a noisy quantum channel. Alice and Bob then make some
measurements on their parts of the EPR pairs and compare results
over a noiseless classical side channel. Based on their
measurements, they then perform local quantum operations to their
remaining qubits to produce a smaller number of more reliable EPR
pairs.  Then, using these EPR pairs and the classical side
channel, Alice teleports her qubits to Bob.  If the EPP has
succeeded, the noise rate in the qubits emerging from the
teleportation protocol is much lower than the noise rate in the
channel.

If we allow Alice to transmit classical information to Bob, but
Bob cannot transmit information to Alice, the EPP is a {\em
one-way} EPP (or 1-EPP). \cite{BDSW} showed that 1-EPPs are
equivalent to QECCs: there is a straightforward procedure to
convert any QECC to a 1-EPP and vice-versa. These protocols are
useful, for instance, to create a quantum memory, in which quantum
information is stored for some time before being used.  The
decoder has to decode the message using only information stored in
the memory, and the encoding cannot depend on information about
the errors which may be gained during decoding. Thus, the
communication is one-way: from the encoder to the decoder.

In contrast, if an EPP is used to transmit information between two
parties, Alice and Bob, there is no reason to prevent Bob from
transmitting classical information to Alice.  An EPP in which
Alice and Bob both transmit classical information is known as a
two-way EPP (or 2-EPP). While a back channel does not help for
transmitting classical data over a classical noisy channel, the
classical back channel {\em does} help in transmitting quantum
data over a noisy quantum channel. 2-EPPs typically tolerate a
much higher error rate than 1-EPPs~\cite{BDSW}, and in some cases
are known to also allow substantially higher data rates even for
low error rates~\cite{erasure,erasure2}.

This channel capacity problem is usually considered in the model
where errors occur independently on different qubits with some
fixed probability, and the goal is to produce a received state
with very high fidelity in the asymptotic limit of many
transmitted qubits.  If we only wish to transmit a few qubits, it
makes more sense to consider a small block code.  For QECCs (and
indeed classical error-correcting codes), we often consider what
is known as the minimum distance scenario, in which we transmit
$n$ qubits to protect $k$ data qubits against up to $t$
single-qubit errors during transmission (or in fact against an
arbitrary error which affects $t$ qubits).  When there are $t$ or
fewer errors, the decoding procedure leaves us with exactly the
correct state on the $k$ data qubits.  When there are more than
$t$ errors, the state can be wrong in arbitrary ways.

We can define the {\em minimum distance problem} for $t$ errors as
follows:
\begin{prob}
Find a protocol which allows Alice to transmit $k$ qubits to Bob
with perfect fidelity. The protocol may use a quantum channel
which transmits $n$ qubits and applies the operation ${\cal S}
\otimes I$, where ${\cal S}$ is some superoperator acting on $t$
of the qubits (not necessarily the first $t$) and $I$ is the
identity on the remaining $n-t$ qubits.  The protocol may use
classical side channels, but cannot depend on any properties of
the quantum channel operation except the fact that it acts
nontrivially on at most $t$ qubits.
\end{prob}
The usual solution to the minimum distance problem is a QECC with
distance $d=2t+1$, which requires no classical side channels.

In this paper, we consider the minimum distance problem for
2-EPPs. Alice prepares $n$ EPR pairs and transmits half of each to
Bob over the noisy quantum channel.  The channel, as described
above, has the property that it applies an arbitrary superoperator
affecting at most $t$ of the pairs. Then the goal of Alice and Bob
is to produce, by talking back and forth over the classical side
channels and performing local operations, at least $k$ good EPR
pairs. Note that, given the promise that there are at most $t$
errors, we insist that the protocol {\em always} works: While some
of the protocols we present will sometimes produce more than $k$
good pairs, we require that they never produce fewer than $k$
correct EPR pairs.  We also require that the EPR pairs produced
are precisely correct (fidelity $1$ to perfect EPR pairs).  A
2-EPP that achieves this will then provide a solution to the
minimum distance problem via quantum teleportation.

Note that for a QECC, the distance encapsulates a number of
properties of the code.  In particular, the distance determines
the number of errors that can be detected and the number of
correctable erasure errors as well as the number of general errors
that can be corrected.  For a 2-EPP, there does not appear to be
any single quantity that encapsulates such a broad set of
properties, so 2-EPPs do not have a ``distance'' in the
conventional sense.

\section{Stabilizer QECCs \& 1-EPPs}

We begin by reviewing the basic theory of stabilizer quantum
error-correcting codes~\cite{intro} and the relationship between
QECCs and 1-EPPs.

\begin{dfn}
The Pauli group $\PG$ is a group consisting of tensor products of
the three matrices
\begin{equation}
X = \begin{pmatrix} 0 & 1 \\ 1 & 0 \end{pmatrix}, Y =
\begin{pmatrix} 0 & -i \\ i & 0 \end{pmatrix}, Z =
\begin{pmatrix}1 & 0 \\ 0 & -1 \end{pmatrix}
\end{equation}
and the identity $I$ with overall phase $\pm 1$, $\pm i$.
\end{dfn}
Note that $X$, $Y$, and $Z$ anticommute with each other (e.g., $XZ
= -ZX$) and that any two elements of the Pauli group either
commute or anticommute.  Furthermore, the Pauli group on $t$
qubits is a basis for the space of all matrices corresponding to
operators acting on $t$ qubits.

\begin{dfn}
A stabilizer $S$ is an Abelian subgroup of $\PG$ which does not
contain $-1$ or $\pm i$.  Let the coding space $C$ be the set of
states $\ket{\psi}$ for which $M \ket{\psi} = \ket{\psi}$ for all
$M \in S$.  Suppose the stabilizer $S$ has $r$ generators $M_1,
\ldots, M_r$. The error syndrome of $P \in \PG$ is the $r$-bit
string whose $i$th bit is $0$ if $P$ commutes with $M_i$ and is
$1$ if $P$ anticommutes with $M_i$.  Let $N(S)$ be the set of
Pauli matrices which have error syndrome $0$ --- i.e., which
commute with the stabilizer.
\end{dfn}

The motivation for the definition of error syndrome and for using
this formalism for defining quantum codes is that if a state
$\ket{\psi}$ is a $+1$-eigenvector of an operator $M$, and $E$
anticommutes with $M$ ($EM = - ME$), then $E \ket{\psi}$ is a
$-1$-eigenvector of $M$.  Thus, looking at a simple property of
the stabilizer allows us to evaluate the code's ability to detect
and correct errors~\cite{Gstab,GF4}.

\begin{thm}
\label{thm:ec}
If there are $n$ qubits and the stabilizer $S$ has $r$ generators,
then the coding space $C$ has dimension $2^{n-r}$.  That is, the
code encodes $k = n-r$ qubits.  The set of undetectable errors for
the code is $N(S) \setminus S$.  The code corrects any set $\E
\subseteq \PG$ for which $E^\dagger F \not\in N(S) \setminus S$
for all $E, F \in \E$.  Thus, the code corrects $t$ errors if
$N(S) \setminus S$ contains no Pauli operations acting on fewer
than $2t+1$ qubits.
\end{thm}

\begin{dfn}
If a stabilizer code corrects a set of errors $\E$, and there
exist $E, F \in \E$ such that $E^\dagger F \in S$ but $E \neq F$,
then the code is said to be {\em degenerate} or {\em impure}.
Otherwise the code is {\em non-degenerate} or {\em pure}.
\end{dfn}

The error correction procedure is simply to measure the eigenvalue
of each of the generators of $S$.  (For each generator, the $2^n$
dimensional space of all states on $n$ qubits decomposes into a
direct sum of two $2^{n-1}$ dimensional subspaces, one consisting
of all eigenvectors of the operator with eigenvalue 1, the other
consisting of all eigenvectors with eigenvalue -1. ``Measuring the
eigenvalue'' means that we measure if the state belongs to the
first or the second of these subspaces, in the process projecting
the state onto the appropriate subspace.)

The correct state has eigenvalue $+1$, but if error $P$ has occurred, the
actual eigenvalue is $-1$.  This gives us the error syndrome, and from
there we can deduce the error and correct it.  For a
non-degenerate code, all the error syndromes are distinct, so the
error syndrome uniquely identifies the error.  The errors are not
uniquely identified for a degenerate code, but it does not matter,
because errors which have the same error syndrome act exactly the
same way on encoded states.

This suggests how we can perform an EPP based on any stabilizer
QECC.  Alice prepares a number of EPR pairs $\ket{00}+\ket{11}$,
and sends the second half to Bob.  If there are no errors in the
channel, Alice, when she measures any Pauli operator $M$ on her
side, will get a predictable measurement result relative to Bob's
when he measures the same $M$ on his side.  In particular, Alice
and Bob can each measure the generators of a stabilizer $S$ on
their own side. In the absence of errors, they should get the same
measurement result for any generator with an even number of $Y$s,
and the opposite measurement result for any generator with an odd
number of $Y$s.%
\footnote{The state $\ket{00}+\ket{11}$ is a $+1$ eigenstate of
$X_A \otimes X_B$ and $Z_A \otimes Z_B$, but a $-1$ eigenstate of
$Y_A \otimes Y_B$.}
That is, if Alice's measurement results form the vector $\bf{a}$
and Bob's measurement results form the vector $\bf{b}$, then
$\bf{a} \oplus \bf{b} = \bf{s}$, where the $j$th bit of $\bf{s}$
is the parity of the number of $Y$s in the $j$th generator of $S$.
On the other hand, if the channel has performed a Pauli error $P$,
they will get different results: $\bf{a} \oplus \bf{b} = \bf{s}
\oplus \bf{e}$, where $\bf{e}$ is precisely the error syndrome of
$P$ with respect to $S$. Thus, if Alice sends her measurement
results to Bob, Bob can compare Alice's results to his, deduce the
error syndrome of $P$, and correct it just as if he were using a
quantum error-correcting code.  Alice and Bob have measured $2r$
qubits, destroying the entanglement of $r$ pairs, but $n-r = k$
pairs are left over.

Note that the remaining entanglement will actually be distributed
across many or all of the original $n$ pairs, so Alice and Bob
must perform a decoding operation to extract it.  Alice and Bob
have each projected their state onto a codeword of the QECC with
stabilizer $S$ with some known syndrome $\bf{a}$ (Alice's
measurement results), and must therefore perform the decoding
operation for $S$ to get back the $k$ EPR pairs they desire.

QECCs and EPPs correct more general errors than just Pauli errors
because of the linearity of quantum mechanics.  In fact, if a code
(or EPP) corrects a set of errors $\cal{E}$, it also corrects any
errors in the linear span of $\cal{E}$.  Therefore, a code or EPP
which corrects Pauli errors on up to $t$ qubits actually corrects
{\em any} error affecting up to $t$ qubits~\cite{intro,shor}, and
thus provides a solution to the minimum distance problem.

\section{Stabilizers and 2-EPPs}

Along the same lines, we can construct a class of 2-EPPs as
adaptive stabilizer codes. The model is as follows. Alice and Bob
measure $r$ commuting Pauli operators, one by one. After measuring
each operator, they both send their measurement results to each
other. Then they XOR the results, obtaining one bit of the error
syndrome. The $(i+1)^{\rm st}$ operator can depend on the results
obtained in the first $i$ measurements but has to commute with all
the previously measured operators. The end result is that Alice
and Bob have measured the generators of one stabilizer code out of
a larger family.  The choice of which code, however, depended on
the results of early measurements. To specify a 2-EPP, we
therefore need to describe a rule for choosing operators to
measure based on the outcomes of previous measurements.  Note that
the procedure is a 2-EPP and not a 1-EPP because Alice needs to
know Bob's results before she knows which operators to measure.
(This type of EPP has been called a {\em stabilizer} EPP, and is
characterized more precisely in Definition 4 of~\cite{twoway}.)

At the end of protocol, Alice and Bob apply local unitary
transformations $U_A$ and $U_B$ which may depend on the results of
the measurements during the protocol. They succeed if, for any
error on at most $t$ EPR pairs, the protocol produces $k$ perfect
EPR pairs.

\subsection{2-EPPs that correct 1 error}
\label{one-error}

As the first example, consider the following 2-EPP which produces
2 good EPR pairs from 6 EPR pairs in the presence of 1 error.
(There is no QECC which can do this for qubits~\cite{GF4}.)  Alice
and Bob measure
\begin{eqnarray}
X \otimes X \otimes X \otimes X \label{eqn:4qubit} \\
Z \otimes Z \otimes Z \otimes Z \nonumber
\end{eqnarray}
on the first four pairs. These two operators generate the
stabilizer for a code detecting an arbitrary single error --- any
single-qubit Pauli operator will be outside $N(S)$.  There are two
possibilities:
\begin{itemize}
\item They detect an error (get a non-zero error syndrome).  In
this case, they know there is an error in the first four pairs,
and therefore none in the last two, since there is a maximum of
one error.  They therefore discard all of the first four pairs,
and keep the remaining two.
\item They detect no error (zero error syndrome).  Since there is
again a maximum of one error, and they would have detected any
single error on the first four pairs, they know the first four
pairs must be correct.  They used up two pairs for the
measurement, but they still have two left. In this case, the two
pairs that are left do not directly correspond to any of the
original four pairs. Instead, Alice and Bob must each perform a
local circuit equal to the decoding operation for the four-qubit
QECC with the appropriate syndrome. If there was no error on the
four pairs at the beginning, then the decoding procedure gives
Alice and Bob two pairs with no errors.
\end{itemize}

It may at first appear that this 2-EPP falls slightly outside the
adaptive stabilizer code construction, as it involves discarding
unwanted pairs.  However, it can easily be rewritten as a
degenerate adaptive stabilizer code, simply by measuring a
complete set of operators for the discarded pairs.  For instance,
if Alice and Bob detect an error with the first two measurements,
they then also measure $X \otimes X \otimes I \otimes I$ and $Z
\otimes Z \otimes I \otimes I$ on the first four pairs.

Next, we present a 2-EPP that produces $2^{m}-m-2$ good EPR pairs
from $2^m-1$ pairs if there is up to one error.  This should be
compared with the best family of 1-error-correcting QECCs known,
which protect $k=2^m-m-2$ qubits with $n=2^m$ qubits if there is
up to one error~\cite{Gstab}.

Alice and Bob start by measuring $X^{\otimes (2^m-1)}$. If they
detect an error, they know that there is a $Y$ or $Z$ error. Then,
they localize it by binary search, in $m-1$ steps. Before the
$i^{\rm th}$ localization step, Alice and Bob have a set $S_i$ of
$2^{m+1-i}-1$ or $2^{m+1-i}$ candidate EPR pairs. They have
measured the product of $X$ on this set and they have detected an
error in it. If $i>1$, they have also measured $i-1$ Pauli
operators on pairs not in $S_i$. In the $i^{\rm th}$ localization
step, Alice and Bob divide $S_i$ into sets $S'_i$ and $S''_i$, one
containing $2^{m-i}$ EPR pairs and the other containing either
$2^{m-i}$ or $2^{m-i}-1$ pairs. Then, they measure the product of
$X$ operators over all pairs in $S'_i$. Together with the previous
measurement (the product of $X$ over $S_i$), this is equivalent to
measuring the product of $X$ in $S'_i$ and the product of $X$ in
$S''_i$. If there is an error in $S'_i$, then Alice and Bob set
$S_{i+1}=S'_i$. Otherwise, $S_{i+1}=S''_i$.

After $m-1$ such steps, Alice and Bob have a set $S_{m}$ consisting of
1 or 2 EPR pairs, they have measured the product of $X$ for $S_m$
and they know that $S_m$ contains a damaged EPR pair.
Then, Alice and Bob discard pairs in $S_m$.
The other $2^m-2$ or $2^m-3$ pairs are
good and only $m-1$ of them have been measured.
(Altogether, Alice and Bob have measured $1+(m-1)$ pairs but one of those
measurements has been on discarded pairs only.)
Therefore, they have at least $2^m-m-2$ good EPR pairs.

The second case is if the first measurement (product of $X$ on all
EPR pairs) detects no error. Then, there is either no error at all
or an $X$ error on one of $2^{m}-1$ EPR pairs. This means that
there are $1+(2^m-1)=2^m$ possibilities for error. With $m$ more
measurements, Alice and Bob can distinguish which one of those has
happened. To do that, they number the EPR pairs by numbers $0, 1,
\ldots, 2^m-2$. Each of those numbers can be written in binary
with $m$ digits. Alice and Bob measure the product of $Z$
operators for all EPR pairs whose numbers have the first digit
equal to 0, the product of $Z$ operators for all EPR pairs whose
numbers have the second digit equal to 0, etc.  That is, they
measure the parity checks for the Hamming code of length
$n=2^m-1$. If there is an $X$ error on the $i^{\rm th}$ EPR pair,
each measurement gives us one bit of $i$ (the bit is 0 if the
corresponding measurement detects an error and 1 if it does not
detect an error). All $m$ measurements together uniquely determine
$i$. It remains to see what happens if there is no error. Then no
measurement detects error, implying that either there is no error
or the error is in the location that has 1 in every bit. If the
location has all $m$ bits equal to 1, its number must be $2^m-1$
but Alice and Bob do not have an EPR pair numbered $2^m-1$.
Therefore, they know for certain that there has been no error.

After that, Alice and Bob know the type of error and its location
and they correct it. They have destroyed $m+1$ EPR pairs.
Therefore, Alice and Bob again have $2^m-m-2$ good EPR pairs
remaining.

We illustrate the protocol with two examples for $m=3$, $2^m-1=7$
and $2^m-m-2=3$. The first is if there is a $Y$ error on the
$3^{\rm rd}$ pair. Then, the operators that Alice and Bob measure
are:
\[ X\otimes X\otimes X\otimes X\otimes X\otimes X\otimes X,\]
\[ X\otimes X\otimes X\otimes X\otimes I\otimes I\otimes I,\]
\[ X\otimes X\otimes I\otimes  I\otimes I\otimes I\otimes I .\]
The first two reveal an error, the third does not. Alice and Bob
conclude that either the $3^{\rm rd}$ or the $4^{\rm th}$ pair has
a $Y$ or $Z$ error. They discard these two pairs. The last two
measurements are equivalent to $X\otimes X\otimes I\otimes
I\otimes I\otimes I\otimes I$ and $I\otimes I\otimes X\otimes
X\otimes I\otimes I\otimes I$. Thus, one measurement has been on
the discarded pairs. Therefore, out of the remaining 5 pairs, 2
have been measured. They have 3 good EPR pairs.

In the second example, there is an $X$ error on the $5^{\rm th}$ pair.
In this case, Alice and Bob measure
\[ X\otimes X\otimes X\otimes X\otimes X\otimes X\otimes X,\]
\[ Z\otimes Z\otimes Z\otimes Z\otimes I\otimes I \otimes I,\]
\[ Z\otimes Z\otimes I\otimes I\otimes Z\otimes Z \otimes I,\]
\[ Z\otimes I\otimes Z\otimes I\otimes Z\otimes I \otimes Z.\]
The first two measurements detect no error. The third and the
fourth detect an error. After that, Alice and Bob know that there
is an $X$ error on the $5^{\rm th}$ pair and correct it. They have
used 4 out of 7 pairs and have 3 good pairs remaining.

\subsection{2-EPPs that correct 2 errors}
\label{two-error}

A more dramatic example is a 2-EPP which produces at least 1 good
EPR pair from 9 EPR pairs when there are up to 2 errors.  This is
better than the quantum Hamming bound, which says that the number
of errors times the number of encoded basis states is at most the
dimension of the overall Hilbert space:
\begin{equation}
2^k \sum_{j=0}^t 3^j \binom{n}{j}  \leq 2^n. \label{eqn:Hamming}
\end{equation}
For $t=2$ errors and $k=1$, this equation would suggest $n \geq
10$.  The six-pair and $2^m-1$ pair 2-EPPs above
also exceed the quantum Hamming
bound for $t=1$. It is not known in general whether the quantum
Hamming bound limits all QECCs, since it could potentially be
violated by a degenerate code, for which distinct errors $E$ and
$F$ have the same error syndrome but act the same way on codewords
(i.e., $E^\dagger F \in S$ for a degenerate stabilizer code).
However, no known QECCs exceed the quantum Hamming bound, and in
fact for $t=2$, linear programming bounds~\cite{lpbound,GF4} show
that $n \geq 11$, so our 2-EPP beats the best QECC by two EPR
pairs.

The particular 2-EPP we present is based on the four-qubit
error-detecting code~(\ref{eqn:4qubit}) and the five-qubit
error-correcting code with stabilizer generators
\begin{eqnarray}
X \otimes Z \otimes Z \otimes X \otimes I \nonumber \\
I \otimes X \otimes Z \otimes Z \otimes X \label{eqn:5qubit}\\
X \otimes I \otimes X \otimes Z \otimes Z \nonumber \\
Z \otimes X \otimes I \otimes X \otimes Z. \nonumber
\end{eqnarray}
The five-qubit code can correct one error or detect two errors.
For the 9-pair EPP, Alice and Bob measure the generators of the
five-qubit code on the first five pairs, and the generators of the
four-qubit code on the last four pairs.  They first use the
results of both measurements to detect errors.  We then have the
following cases:
\begin{itemize}

\item They detect an error on the last four pairs.  In that case,
there can be at most one error on the first five pairs, so they
can use the five-qubit code to correct the error, producing one
good pair. They discard the last four pairs, leaving them with one
good pair overall.

\item They detect an error on the first five pairs but none on the
last four pairs.  In that case, there is at least one error on the
first five pairs, so there could be at most one on the last four.
If there had been one error on the last four, they would have
detected it, so Alice and Bob know the last four pairs are safe.
They discard the first five pairs (which could contain two
errors), and extract the two remaining pairs from the last four.
In this case, they are left with two good pairs.

\item They detect no errors on either set of pairs.  If there had
been any errors (one or two) on the first five pairs, they would
have detected them.  Therefore, any errors must be on the last
four pairs.  It is possible, however, that two errors on the last
four pairs would go undetected.  They discard the last four pairs,
and extract the one remaining pair from the first five pairs;
there is no need for error correction.  They are left with one
good pair.

\end{itemize}

\subsection{2-EPPs in higher dimensions}

We can also create EPPs that violate the quantum Singleton
bound~\cite{KL}
\begin{equation}
n \geq 4t + k. \label{eqn:Singleton}
\end{equation}
This bound applies to degenerate quantum codes as well as
nondegenerate ones, and shows, for instance, that the smallest
QECC to correct one error has 5 qubits.  The construction we
present requires using registers with dimension greater than two;
qutrits will suffice.  The quantum Singleton bound also applies to
higher-dimensional codes, so there is no QECC encoding 1 qutrit in
4 and correcting one error.  In contrast, we present a 2-EPP that
corrects one error out of only 4 pairs.

We can generalize the stabilizer formalism to a higher dimension
$d$ by replacing the qubit Pauli group with the group generated by
tensor products of $X: \ket{j} \rightarrow \ket{j+1}$ and $Z:
\ket{j} \rightarrow \omega^j \ket{j}$, where addition is modulo
$d$, and $\omega = \exp(2\pi i/d)$~\cite{Kqudit}.  The eigenvalues
of elements of this higher-dimensional Pauli group are powers of
$\omega$, and $PQ = \omega^{r(P,Q)} QP$, where $P$ and $Q$ are
arbitrary elements of the Pauli group and $r(P,Q)$ is an integer
function of $P$ and $Q$.  The same basic principle allows us to
create stabilizer codes in higher dimensions: if $M \ket{\psi} =
\ket{\psi}$ and $MP = \omega^r PM$, then $M\,(P \ket{\psi}) =
\omega^r P \ket{\psi}$.  We can therefore again create codes as
the joint $+1$-eigenspace of elements of an Abelian subgroup of
the Pauli group, and the Pauli errors it detects will again be all
operators outside $N(S) \setminus S$.

In particular, we can define a 3-qutrit QECC to {\em detect} one
error using the stabilizer
\begin{eqnarray}
X \otimes X \otimes X \label{eqn:3qutrit} \\
Z \otimes Z \otimes Z. \nonumber
\end{eqnarray}
Similar codes exist for many larger-dimensional registers as well.%
\footnote{In particular, Reed-Solomon codes can be used to
construct a code with these parameters over any finite field GF(q)
with $q=p^s > 2$.  Then, treating prime power factors of a
register's dimensionality separately allows us to construct an
appropriate QECC for any dimension which is odd or a multiple of
4, leaving open the cases where the dimension is 2(2k+1).}
We can then create a 2-EPP correcting 1 error out of 4 qutrit EPR
pairs.  Alice and Bob measure the generators of this
error-detecting code on the first three pairs.  If they detect an
error, they keep the fourth pair and discard the first three.
Otherwise, they discard the last pair and extract the one
remaining pair from the first three.  Either way, they end up with
one reliable EPR pair out of the original four.

\section{Asymptotic Lower Bound}

To construct the above examples of 2-EPPs, we split up the EPR
pairs and used error detection techniques to discard noisy pairs.
This does not work well for protocols with many pairs and
proportionally many errors, but EPPs can still do substantially
better than QECCs in the asymptotic regime.

\begin{thm}
\label{thm2} For all $n$, for any set of errors $\E \subseteq
\PG$, there exist 2-EPPs producing $k$ EPR pairs from $n$ pairs
correcting $\E$ satisfying
\begin{equation}
k \geq n - \log_2 |\E| - 2.
\end{equation}
\end{thm}

\begin{cor}
\label{cor} For all $n$ and $t$, there exist 2-EPPs producing $k$
EPR pairs from $n$ pairs with up to $t$ errors satisfying
\begin{equation}
k \geq n - \log_2 \left[ \sum_{j=0}^t 3^j \binom{n}{j} \right] -
2.
\end{equation}
\end{cor}

That is, 2-EPPs can come within two qubits of the quantum Hamming
bound for all values of $n$ and $t$.  This is in contrast to the
case for QECCs, for which the best general lower bound is the
quantum Gilbert-Varshamov bound~\cite{CRSS}, which shows that
there exist stabilizer codes satisfying
\begin{equation}
k \geq n - \log_2 \left[\sum_{j=0}^{2t} 3^j \binom{n}{j} \right].
\end{equation}
(The sum is taken to $2t$ rather than $t$.)  In fact, the lower
bound from the corollary is actually better in many cases than the
general {\em upper} bounds proved on QECCs via linear
programming~\cite{lpbound,ALbound}.

\begin{figure}[th]
\epsfxsize=3in \hspace{0in} \epsfbox{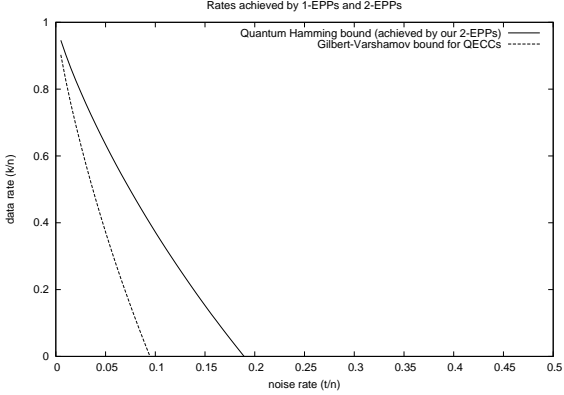} \vspace*{8pt}
\caption{Lower bounds on rates achieved by 1-EPPs and 2-EPPs}
\label{fig:rates}
\end{figure}

\begin{proof}
To prove Theorem \ref{thm2}, we have Alice and Bob build up their
stabilizer $S$ element by element.  At each stage, there is a set
$\E$ of possible errors compatible with all available information.
Initially, for instance, $\E$ may be the set of all Pauli errors
of weight up to $t$ (as in the corollary), and later $\E$ would be
the set of Pauli errors of weight up to $t$ which have a
particular error syndrome relative to the current stabilizer $S$.
As Alice and Bob add more generators to their stabilizer, the set
$\E$ shrinks. Once they have narrowed $\E$ down to a single error,
they can correct the state.  If they have measured $r$ generators
at this point, they have $n-r$ EPR pairs remaining after decoding.
Thus, the goal is to show that Alice and Bob can reduce the size
of $\E$ to $1$ by measuring at most $n-k$ stabilizer generators.
It is sufficient to consider the case where $\E$ does not include
two Pauli operators differing only by a factor of $-1$ or $\pm i$.

Suppose we are somewhere in the middle of this procedure, with the
set $\E$ of currently possible errors. Alice and Bob now must
choose a new generator $M$ to measure and add to $S$. $M$ must
commute with everything in $S$, of course, and should be
independent of the previous elements of $S$ (i.e., $M \in N(S)
\setminus S$). There are many possible $M$s, but Alice and Bob
wish to choose one that comes as close as possible to dividing the
set of possible errors in half.  That is, $M$ commutes with close
to half of $\E$ and anticommutes with close to half of $\E$.

Let $C(M) \subseteq \E$ be the set of possible errors that commute
with $M$ and $A(M) \subseteq \E$ be the set of possible errors
that anticommute with $M$. Then, when Alice and Bob measure $M$,
the new set of possible errors will be either $C(M)$ or $A(M)$,
depending on the measurement result.  In the worst case, it will
be the larger of these two sets, so our goal is to show that
$\max(|C(M)|, |A(M)|)$ is not much larger than $|\E|/2$.  Alice
and Bob repeat this process until the set of possible errors has
shrunk to a single operator, at which point they know the error
and can correct it. The number of generators they must add to the
stabilizer to do this is $n-k$, and we wish to show that in the
worst case, $n-k$ is not much larger than
\begin{equation}
\log_2 |\E| = \log_2 \left[ \sum_{j=0}^t 3^j \binom{n}{j} \right].
\end{equation}

For any $E, F \in \E$, we say $P \in N(S)$ {\em separates} the pair $(E,
F)$ iff $E^\dagger F \in A(P)$, so $E \in A(P)$ and $F \in C(P)$
or vice-versa.  In fact, precisely half of the elements $P \in
N(S)$ separate any pair $(E,F)$, but no element of $S$ does (since
$E$ and $F$ have the same error syndrome relative to $S$). If $|S|
= 2^r$, then $|N(S)| = 2^{n-r}$, so each pair $(E,F)$ is separated
by $2^{n-r-1}$ elements of $N(S) \setminus S$.  There are
$\binom{|\E|}{2}$ pairs total, so collectively, the elements of
$N(S) \setminus S$ separate $2^{n-r-1} \binom{|\E|}{2}$ pairs. On
average, then, the elements of $N(S) \setminus S$ separate
\begin{equation}
\frac{2^{n-r-1}}{2^{n-r} - 2^r} \binom{|\E|}{2} > \frac{1}{2}
\binom{|\E|}{2}
\end{equation}
pairs each.  In particular, there exists $M \in N(S) \setminus S$
that separates at least this many pairs.

Now, $M$ has the sets $C(M)$, $A(M)$ and separates $|C(M)| \cdot
|A(M)|$ pairs.  Also note $|C(M)| + |A(M)| = |\E|$.  Thus,
\begin{equation}
m (|\E| - m) > \frac{1}{2} \binom{|\E|}{2},
\end{equation}
where $m = \max(|C(M)|, |A(M)|)$.  For instance, when $|\E| = 4$,
we find $m (4-m) > 3$, and since $m$ is an integer, $m=2$.

If we set $m = |\E|/2 + \epsilon$, then we find
\begin{equation}
\frac{1}{4} |\E|^2 - \epsilon^2 > \frac{1}{4} (|\E|^2 - |\E|),
\end{equation}
or
\begin{equation}
\epsilon^2 < |\E|/4. \label{eqn:diff}
\end{equation}

Using (\ref{eqn:diff}) repeatedly, and bearing in mind that $m$
must always be an integer, we can find the number of steps
necessary to bring any particular initial value of $|\E|$ down to
$1$.  At any stage, given $\E$, choosing another stabilizer
generator by the above rule gives us a new set $\E'$ of possible
errors, with
\begin{equation}
|\E'| < (|\E| + \sqrt{|\E|})/2.
\end{equation}

We can define an integer sequence $m_i$ such that $m_0 = 1$ and
$m_i$ is the largest integer such that
\begin{equation}
(m_i + \sqrt{m_i})/2 \leq m_{i-1}+1. \label{eq:mi}
\end{equation}
Thus, whenever $|\E| \leq m_i$, it follows that $|\E'| \leq
m_{i-1}$. We can therefore reduce the set of possible errors to
$1$ in at most $i$ steps. Below we give $m_i$ for small values of
$i$:
\begin{equation}
1 \la 2 \la 4 \la 7 \la 12 \la 21 \la 37 \la 67 \la 124 \la 234.
\end{equation}

For larger values of $i$, we note that
\begin{eqnarray}
2m_i \, \geq \, m_{i+1} \!\!\! & \geq & \!\!\! \lfloor 2(m_{i}+1)
- \sqrt{2(m_i+1)}
\rfloor \\
& \geq & \!\!\! 2m_i - \sqrt{2m_i} - 1.
\end{eqnarray}
If
\begin{equation}
2^{i-1} \geq m_{i} \geq 2^{i-2} + \delta_i,
\end{equation}
then
\begin{equation}
2^{i} \geq m_{i+1} \geq 2^{i-1} + 2\delta_i - \sqrt{2^i} - 1.
\end{equation}
Let $\delta_{i+1} = 2\delta_i - 2^{i/2} - 1$, and let $\delta_8 =
60$ (since $2^7 \geq m_8 = 124 \geq 2^6 + 60$).  Then for $i \geq
8$, $\delta_i \geq (1.5)^{i-8} 60$ by induction:
\begin{equation}
\delta_{i+1} \geq (1.5)^{i-7} 60 + \left[ (1.5)^{i-8} 30 -
(\sqrt{2})^{i-8} 16 - 1 \right].
\end{equation}
In particular, $\delta_i \geq 0$ for $i \geq 8$, so for $i \geq
8$, $2^{i-1} \geq m_i \geq 2^{i-2}$.  Therefore, when $|\E| \leq
2^j$, we can reduce the set of possible errors to $1$ in at most
$j+2$ steps, proving the theorem.

\end{proof}

Note that this technique of narrowing down the set of possible
errors fails if we try to apply it to QECCs. While we can indeed
choose a single stabilizer generator $M_1$ which separates $\E$
into approximately equal sets $A(M_1)$, $C(M_1)$, choosing the
second generator $M_2$ is more difficult.  For 2-EPPs, we need
only consider one of the two sets $A(M_1)$, $C(M_1)$, whichever is
indicated by the first measurement.  For a QECC, we do not know
which set will be selected, so the second generator $M_2$ must
divide both of these sets approximately in half.  The problem
compounds at later steps, as the third generator chosen must
simultaneously divide up four sets of possible errors, and the
$i^{\rm th}$ generator must be chosen to evenly divide up
$2^{i-1}$ different sets of possible errors all at once.  Clearly
this is substantially more difficult than splitting just a single
set in half, and results in a significantly reduced efficiency for
the QECC compared to a 2-EPP.

\section{Conclusion}

We have considered the minimum distance scenario for 2-EPPs and
given a number of examples of 2-EPPs that are more efficient than
any QECC.  The small block EPPs we present might be useful for
quantum communication in near-future scenarios.  The asymptotic
construction we give of 2-EPPs is not very practical, since
finding the optimal set of measurements appears to be a
computationally difficult task.  For practical applications of
2-EPPs, we want the equivalent of an efficient decoding algorithm
--- namely, an efficient algorithm to tell us what to measure
next, and, once all measurements are complete, to tell us how to
correct the state.

One way to find such EPPs might be to consider QECCs with good
list-decoding algorithms.  Since classical error-correcting codes
can be substantially more efficient when we only demand list
decoding rather than minimum distance decoding~\cite{sudan}, it
seems very likely that QECCs would have the same property. Then we
might be able to convert the list-decoded QECC to a
minimum-distance 2-EPP by choosing just a few additional
generators to narrow down the short list of possible errors to a
single error.

\section*{Acknowledgements}

A.~A. is supported by CIAR, NSERC and IQC University
Professorship.  D.~G. is supported by CIAR and NSERC.  Part of
this work was done while A.~A. was at the Institute for Advanced
Study (supported by NSF Grant CCR-9987845 and the State of New
Jersey) and D.~G.~was a Long-Term Prize Fellow for the Clay
Mathematics Institute. A.~A. thanks Ke Yang for discussions that
brought this problem to his attention.

%
%
%

\end{document}